\documentclass[ reprint,
amsmath,amssymb,
aps,
]{revtex4-2}

\usepackage{tikz}
\usetikzlibrary{arrows,snakes,backgrounds}

\usepackage{epsf,epsfig}
\usepackage{amssymb,amsmath,amsfonts}
\usepackage{color}
\usepackage{braket}
\usepackage{subcaption}
\usepackage[T1]{fontenc} 

\usepackage{physics}
\usepackage{tikz}
\usetikzlibrary{arrows,snakes,backgrounds}

\usepackage[english]{babel}

\usepackage{graphicx}

\usepackage{xcolor}
\usepackage[colorlinks=true,
            linkcolor=magenta,
            citecolor=blue,
            urlcolor = magenta]{hyperref}

\definecolor{NordCyan}{HTML}{8FBCBB}          
\definecolor{NordBrightCyan}{HTML}{88C0D0}    
\definecolor{NordBlue}{HTML}{81A1C1}          
\definecolor{NordBrightBlue}{HTML}{5E81AC}    
\definecolor{NordRed}{HTML}{BF616A}           
\definecolor{NordOrange}{HTML}{D08770}        
\definecolor{NordYellow}{HTML}{EBCB8B}        
\definecolor{NordGreen}{HTML}{A3BE8C}         
\definecolor{NordMagenta}{HTML}{B48EAD}       

\begin{document}

\title{Reconstruction of Black Hole Islands from Hawking Radiation\\ via the Algebraic Petz Map
}


\author{Niloofar Vardian}
\email{vardian@sharif.edu}
\affiliation{Department of Physics, Sharif University of Technology,\\
P.O.Box 11155-9161, Tehran, Iran}
\affiliation{Research Center for High Energy Physics, Department of Physics, Sharif University of Technology,\\
P.O.Box 11155-9161, Tehran, Iran
}

\begin{abstract}
\noindent

The island proposal guarantees a unitary Page curve for evaporating black holes, implying that operators deep behind the horizon must be fully reconstructable from asymptotic Hawking radiation. However, because the interior is causally disconnected and highly dynamical, standard entanglement wedge reconstruction methods—relying on global isometries or continuous modular flow—fail to provide this mapping. We resolve this by constructing a protocol for island reconstruction using the algebraic Petz map. By using modular theory, we reflect island operators into the accessible complement, transport them to the boundary, and map them into the radiation algebra. Furthermore, we demonstrate that this purely algebraic derivation naturally yields the structure of Papadodimas-Raju mirror operators. This framework provides a rigorous operational dictionary for the island formula, proving that information trapped inside a black hole can be decoded.

\end{abstract}
\maketitle
\section{Introduction}

The island formula provides a striking resolution of the black-hole
information paradox at the level of entropy: in an evaporating black hole coupled to a non-gravitating bath after the Page time, the
entanglement wedge of the Hawking radiation is a bulk island,
yielding a unitary Page curve~\cite{Penington:2019npb,Almheiri:2019psf,page1993information}.
An operational understanding of this result, however, requires more
than the entropy formula. One should exhibit a reconstruction map that
takes an operator in the island algebra and produces an operator acting
only on the asymptotic Hawking radiation. Such a map is subtle because
the island is not, in general, directly accessible through the
bulk-to-boundary reconstruction map used for ordinary entanglement
wedge reconstruction (EWR).

A \emph{general} EWR at leading order can be formulated either through
modular flow by Faulkner and Lewkowycz \cite{Faulkner:2017} or through quantum error correction~\cite{
Cotler:2017erl, Chen:2019gbt, Bahiru:2022ukn}. In the
semiclassical regime, modular-flow reconstruction expresses bulk
operators in an entanglement wedge in terms of boundary operators,
while the quantum-error-correction formulation associates reconstruction
with a recovery channel. For an evaporating black hole, neither
viewpoint immediately provides a direct island-to-radiation map: the
island is separated from the asymptotic radiation by a dynamical
horizon
\footnote{The modular-flow approach relies crucially on the JLMS relation \cite{jafferis2016relative} (which can be applied to the case of islands), together with the modular-evolved extrapolate dictionary, whose applicability is not straightforward in settings involving islands.}, while a reconstruction channel for the island algebra itself
is not part of the usual bulk-to-boundary construction. The relevant
question is therefore not how to construct a direct map from the island
to the boundary, but whether the island algebra can be related
algebraically to a complementary algebra for which such a map is
available.

Here we show that this can be achieved using Tomita--Takesaki modular
theory. Let $\mathcal A_{\mathcal{I}}$ denote the algebra of island observables and
$\mathcal A_{\bar {\mathcal{I}}}$ its algebraic complement within the semiclassical
code algebra. Assuming that the relevant state is cyclic and separating
and that the appropriate algebraic duality relation $     \mathcal A_{\mathcal{I}}' = \mathcal A_{\bar {\mathcal{I}}}$
holds, modular conjugation provides an anti-isomorphism
$    J_{\mathcal{I}}\,\mathcal A_{\mathcal{I}}\,J_{\mathcal{I}} = \mathcal A_{\bar {\mathcal{I}}}.$
The essential observation is that the complementary algebra
$\mathcal A_{\bar {\mathcal{I}}}$ lies in the entanglement wedge accessible to the
asymptotic CFT after the Page time. Consequently, the semiclassical
EWR quantum channel
\begin{equation}
    \mathcal {N}_{\bar{\mathcal{I}}}:
    \mathcal A_{\bar {\mathcal{I}}}
    \longrightarrow
    \mathcal A_{\rm CFT}
\end{equation}
can be used even though no corresponding channel for
$\mathcal A_{\mathcal{I}}$ is known.

The appropriate action on the commutant is the
state-dependent algebraic $\psi$-dual map $(\mathcal N^*)'_\psi$~\cite{furuya2020real}. Combining this dual map with modular
conjugation gives a reconstruction map from the island algebra directly
into the radiation algebra.
For $\phi_{\mathcal{I}}\in\mathcal A_{\mathcal{I}}$, this gives
\begin{equation}
\boxed{
   \Phi_{\mathcal{R}} = J_{\text{bdy}} \; (\mathcal N^*)'_\psi\!\left( J_{\text{bulk}} \, \phi_{\mathcal{I}} \, J_{\text{bulk}} \right) J_{\text{bdy}}~ \in ~ \mathcal{A}_{\mathcal{R}}
}
\label{eq:main_operator_intro}
\end{equation}
The three factors have a direct interpretation: $J_{bulk}$ moves the island
operator to its accessible commutant, $(\mathcal N^*)'_\psi$ transports the
resulting operator to the boundary commutant (commutant of the radiation), and $J_{bdy}$ maps it into the
physical radiation algebra. Thus the construction does not require a
bulk-to-boundary reconstruction map whose domain contains the island
itself. Instead, the missing reconstruction channel is bypassed through
the commutant.

For an evaporating black hole, we apply this construction to the
post-Page-time geometry, taking $\mathcal A_{\bar {\mathcal {I}}}$ to be the algebra
associated with the complementary entanglement wedge of the asymptotic
CFT. The required boundary channel is then supplied by Faulkner and Lewkowycz work~\cite{Faulkner:2017}.
Within the semiclassical regime, the above composition gives an
 algebraic dictionary between island operators and
Hawking-radiation operators. The resulting radiation operators are
intrinsically state dependent because both modular conjugations and the
dual map are defined relative to the reference state. In particular,
the structure of the corresponding island mirror operators follows
directly from modular theory rather than being introduced as an
independent prescription, in the spirit of the state-dependent mirror
operators of Papadodimas and Raju~\cite{Papadodimas:2012aq, Papadodimas:2013jku, Papadodimas:2015jra}.

Our construction should be distinguished from a direct application of
the ordinary Petz map~\cite{Cotler:2017erl,Chen:2019gbt}. The latter requires a
reconstruction channel whose domain contains the algebra to be
recovered, whereas here the channel is defined on the complementary
algebra and the modular conjugation transfers the reconstruction
problem to that domain. The resulting map therefore isolates the
algebraic mechanism by which the island formula can be promoted from an
entropic statement to an operator-level reconstruction statement.

The construction is algebraic, while its holographic input is
semiclassical. In particular, the use of modular-flow EWR and the identification of the relevant bulk channel are
understood at the order in the semiclassical expansion at which those
results hold. For local quantum-field-theory algebras, which are
generically type III, density-matrix and trace expressions that appear
in regulated representations should likewise be understood as
shorthand; the reconstruction map itself is formulated in terms of von
Neumann algebras, modular conjugations, and the state-dependent dual
channel. Within these assumptions, the result provides an explicit
route from the island algebra to the algebra of Hawking radiation and
gives an algebraic origin for the mirror operators associated with
black-hole islands.

\section{Entanglement Wedge Reconstruction: Two Parallel Approaches}

Because the island $\mathcal{I}$ dynamically enters the entanglement wedge of the radiation $\mathcal{R}$ after the Page time, EWR dictates that any island operator $\phi_{\mathcal{I}}$ must be reconstructable from $\mathcal{R}$ alone. Traditionally, a general EWR relies on two powerful approaches, both of which face fundamental barriers for evaporating black holes.

The first approach, by Faulkner and Lewkowycz \cite{Faulkner:2017}, utilizes \emph{modular flow}. For a boundary subregion $\mathcal{R}$, the modular Hamiltonian $K_{\mathcal{R}}$ generates an algebraic flow on local operators, $\mathcal{O}_s(x) = e^{i K_{\mathcal{R}} s} \mathcal{O}(x) e^{-i K_{\mathcal{R}} s}$. Via the JLMS relation \cite{jafferis2016relative}, $K_{\mathcal{R}}$ acts in the bulk as the modular Hamiltonian $K_{\text{bulk}, r}$, driving a flow inside the entanglement wedge. This allows reconstruction deep inside the causal shadow:
\begin{equation}
\phi(X) = \int_{\mathcal{R}} dx \int_{-\infty}^{\infty} ds \, f_s(X|x) \mathcal{O}_s(x) + \mathcal{O}(1/N),
\end{equation}
where $f_s(X|x)$ is a smearing kernel.

However, the Faulkner--Lewkowycz construction uses not only
modular flow but also the modular-evolved extrapolate
dictionary, which relates bulk fields near the asymptotic
AdS boundary to boundary operators. Its direct application
to an island is therefore not straightforward: the island
is disconnected from the asymptotic region associated with
the radiation, and the corresponding extrapolate-dictionary
input is not available in the same form. Consequently, the
original Faulkner--Lewkowycz prescription does not directly
provide the island-to-radiation reconstruction considered
here.
This limitation concerns that particular construction,
rather than modular flow as a general reconstruction tool.
Indeed, Chen \cite{Chen:2019iro} developed an island
reconstruction using the microscopic radiation modular
Hamiltonian, with an explicit realization for free massless
fermions in JT gravity coupled to baths. That construction
does not amount to a direct application of the original
Faulkner--Lewkowycz smearing prescription to a disconnected
island.

The second approach frames EWR as quantum error correction \cite{Almheiri:2014lwa}. Mapping the bulk state to $\mathcal{R}$ defines a noisy quantum channel. Because JLMS guarantees the preservation of relative entropy between bulk and boundary, this channel is perfectly reversible. Bulk operators are therefore reconstructable via the Heisenberg-picture \emph{Petz map} \cite{cotler2019entanglement, chen2020entanglement}.
While this avoids the need for flows, its standard application is critically limited: constructing 
the channel
typically requires a global bulk-to-boundary isometric embedding (e.g., global HKLL  \cite{Bena:1999jv, hamilton2006local, hamilton2006holographic, hamilton2007local, hamilton2008local}). In an evaporating black hole, the CFT causal wedge covers only the exterior; the island lies completely outside the domain of any known bulk-to-boundary isometry (HKLL reconstructions).

Thus, neither modular flow by Faulkner and Lewkowycz  nor the ordinary Petz map can directly recover island operators from the Hawking radiation.

\section{Island Reconstruction}

In the previous section, we mentioned that  the standard Petz recovery map using a global isometry obtained from the maps like the global  HKLL. This works when the bulk geometry has a global Cauchy slice that can be fully mapped to the boundary — for example, pure AdS or asymptotically AdS spacetimes without horizons.

However, when a black hole is present, no such global isometry is known. The causal wedge of the entire CFT covers only the exterior of the black hole. The interior region — and the island — cannot be accessed by standard HKLL. Therefore, we need a more general reconstruction method that does not rely on a global embedding isometry 
\footnote{Explicit horizon-crossing reconstruction has also been developed
for probe scalar fields in eternal two-sided AdS black holes,
using maximal slicing and operators from both boundary CFTs,
for example look at \cite{Kaushal:2025wbn}. This differs from the radiation-only
reconstruction of island observables considered here.}.

This generalisation is provided by the \emph{algebraic Petz map} introduced in  Appendix \ref{A}, which uses only the algebraic structure of local operators and the concept of  \emph{modular conjugation} from Tomita–Takesaki theory. The key idea is to replace the missing isometry with modular data that exist for any state that is cyclic and separating.

For a boundary region $A$ and its bulk entanglement wedge $a$, we wish to reconstruct a bulk operator $\phi_a \in \mathcal{A}_a$ as a boundary operator $\Phi_A \in \mathcal{A}_A$. The algebraic Petz map achieves this by exploiting the commutant algebras (the spatial complements $\bar{a}$ and $\bar{A}$). Assuming a cyclic and separating state, the explicit reconstruction formula is given by:
\begin{equation} \label{eq:general_algebraic_petz}
    \Phi_A = J_{\text{bdy}} \, (\mathcal N^*)'_\psi \big( J_{\text{bulk}} \, \phi_a \, J_{\text{bulk}} \big) \, J_{\text{bdy}}.
\end{equation}
This equation operates in three distinct steps:
\begin{itemize}
    \item  $J_{\text{bulk}}$ reflects the bulk operator into the complementary region $\bar{a}$ \footnote{Note that this definition works when $\bar{a}= \mathcal{E}_{\bar{A}}$, look at \cite{Hayden:2018khn} for other possibilities};
    \item the $\psi$-dual map 
    $(\mathcal N^*)'_\psi$ (acting purely on the commutants) transports it to the boundary complement $\bar{A}$; and 
    \item  the boundary modular conjugation $J_{\text{bdy}}$ reflects it back into the physical boundary region $A$.
\end{itemize}
Note that, mostly, the $\psi$-dual map can be written explicitly in the cases that $\bar{a} = \mathcal{E}_{\bar{A}}$. As an example, look at \cite{Hayden:2018khn} for the other possible scenarios.


We now apply this framework directly to the island proposal. We consider a Cauchy slice $\Sigma$ after the Page time and partition it as
$
\Sigma= \mathcal{I}\cup\bar{\mathcal{ I}},
$
where $\mathcal{I}$ denotes the island and $\bar{\mathcal{I}}=\Sigma\setminus \mathcal{I}$ its complement. Let $\mathcal A_{\mathcal{I}}$ and $\mathcal A_{\bar{\mathcal{I}}}$ denote the corresponding von Neumann algebras within the semiclassical code algebra. We assume that, within this code algebra, the island and its complement satisfy the appropriate algebraic duality relation
\begin{equation}\label{haag}
\mathcal A_{\mathcal{I}}'=\mathcal A_{\bar {\mathcal{I}}}.
\end{equation}
This relation is an assumption about the algebras and code subspace under consideration, rather than a general consequence of spacelike separation. Under this assumption, and for a cyclic and separating state, Tomita--Takesaki modular conjugation provides an anti-isomorphism between the two algebras,
\begin{equation} \label{Haag}
J_{\rm bulk}~\mathcal A_{\mathcal{I}}~J_{\rm bulk}=\mathcal A_{\bar {\mathcal{I}}}.
\end{equation}
The reconstruction below therefore relies on the existence of this algebraic complementarity relation within the chosen semiclassical code algebra.
For concreteness, we take the bulk matter sector to be a free scalar field and assume the standard Haag-duality property for the corresponding local von Neumann algebras. Thus, for complementary regions $\mathcal{I}$ and $\bar {\mathcal{I}}$ on the chosen Cauchy slice, \eqref{haag}, and \eqref{Haag} hold.
Note that in gauge theories and especially gravity, Haag duality is not automatic and gravitational dressing can make the algebraic situation even more subtle.
More generally, the construction below requires the corresponding algebraic duality relation within the semiclassical code algebra.

It is worth noting that gravitational dressing does not always obstruct the present reconstruction. Following the state-dressing construction \cite{Bahiru:2022oas, Bahiru:2023zlc}, and its refinement for entanglement islands in~\cite{antonini2025apologia}, one can construct gauge-invariant dressed operators that are appropriately localized in the island. The algebraic Petz construction can therefore be applied directly to these dressed observables. Other relevant works include ~\cite{Geng:2024dbl, Geng:2025rov, Geng:2025gns, Geng:2025bcb, Geng:2025byh, Geng:2025gqu, Geng:2026asi, elsayed2026observablealgebrajtdesitter}.

The island lies in the entanglement wedge of the radiation: $\mathcal{I} = \mathcal{E}_\mathcal{R}$. Its complement lies in the entanglement wedge of the entire CFT: $\bar{\mathcal{I}} = \mathcal{E}_{\text{CFT}}$. There is no known direct map from $\mathcal{I}$ to the boundary (neither to $\mathcal{R}$ nor to the CFT).

More formally, let us consider the abstract Schrodinger picture quantum channels 
\(\mathcal{N}_{\mathcal{I}} : \mathcal{A}_{\mathcal{I}} \to \mathcal{A}_{\mathcal{R}}\) 
and \(\mathcal{N}_{\bar{\mathcal{I}}} : \mathcal{A}_{\bar{\mathcal{I}}} \to \mathcal{A}_{\text{CFT}}\)
\footnote{We use the same notation for the
Schrodinger-picture channel acting on density operators
and its linear extension to the full operator space in
a finite-dimensional regulated description. Accordingly,
the displayed algebra-to-algebra notation refers to this
linear extension, while the physical channel acts on
density operators. This convention does not identify
the map with its Heisenberg adjoint, which acts on
observables in the opposite direction.}.
From the JLMS statement \cite{jafferis2016relative}, we know that both are reversible on the corresponding set of density matrices, and the recovery channel can be given via the Petz recovery channels \cite{Cotler:2017erl}.
Because the state of the CFT after the Page time can be well-approximated by a thermal state, 
the bulk and boundary reference states satisfy the cyclic and separating conditions. 
Following the framework developed in \cite{Vardian:2023fce} for static black holes (both eternal and one-sided black holes), 
one can construct the \(\psi\)-dual map \((\mathcal{N}_{\mathcal I}^*)'_\psi\) from the complement channel 
\(\mathcal{N}_{\bar{\mathcal{I}}}\). In the matrix algebra shorthand, for an operator $ \phi_{\bar{\mathcal I}} \in \mathcal{A}_{\bar{\mathcal I}}$, we have 
\begin{equation}
    (\mathcal{N}_{\mathcal{I}}^*)'_\psi ( \phi_{\bar{\mathcal I}}) = \rho_{CFT}^{-1} ~\mathcal{N}_{\bar{\mathcal{I}}}\big( \rho_{\mathcal{\bar{I}}} ~ \phi_{\bar{\mathcal I}} \big).
\end{equation}
We shall denote \( (\mathcal{N}_{\mathcal{I}}^*)'_\psi\) 
simply as \( (\mathcal{N}^*)'_\psi\) from now on.


Although the QES and the associated island algebra evolve during
Hawking evaporation, we work in a quasi-static regime, away from
transitions between competing QES saddles. 

At each time $t$, we choose compatible faithful reference states
on the bulk and boundary algebras, represented by cyclic and
separating vectors in their standard representations. The
state-dependent dual map $(\mathcal{N}_t^*)'_{\psi_t}$ is then
defined with respect to these instantaneous reference states.
At each time $t$, we work with an instantaneous semiclassical
code subspace centered on the time-evolved bulk reference state
$|\psi_{\mathrm{bulk}}(t)\rangle$, together with its corresponding
boundary code subspace centered on
$|\Psi_{\mathrm{bdy}}(t)\rangle$. The bulk code subspace is
generated by a restricted set of low-energy excitations about
the reference state, chosen so that their backreaction does
not significantly alter the background geometry or the
relevant QES configuration. The bulk and boundary reference
states, and the associated reconstruction data, are assumed
to be compatible under the holographic correspondence.
The modular conjugations and the state-dependent dual map
are defined relative to these reference states at each time.  From now on, we suppress the subscript $t$. 

The core of our proposal rests on a simple but powerful physical observation: although the island $\mathcal{I}$ itself has no direct map to the boundary, its complement $\bar{\mathcal{I}}$ can lie entirely within the accessible entanglement wedge of the total asymptotic system. Consequently, operators in $\bar{\mathcal{I}}$ \emph{can} be explicitly mapped to the boundary using established methods. Specifically, the Faulkner-Lewkowycz modular flow construction provides a well-defined embedding isometry for this region, which naturally defines a valid quantum channel in the Schrodinger picture $\mathcal{N}_{\bar{\mathcal{I}}} : \mathcal{A}_{\bar{\mathcal{I}}} \to \mathcal{A}_{\text{CFT}}$ mapping the bulk complement to the boundary degrees of freedom.

With this channel in hand, we can universally reconstruct any operator $\phi_{\mathcal{I}} \in \mathcal{A}_{\mathcal{I}}$ trapped inside the island by substituting our operator into the general algebraic Petz formula \eqref{eq:general_algebraic_petz}.

To be more precise, at one fixed post-Page time, denote a cyclic-separating bulk state by  $|\psi_{\text{bulk}}\rangle$. For the bulk state $|\psi_{\text{bulk}}\rangle$, the modular conjugation associated with the pair $(\mathcal{A}_{\mathcal{I}}, \mathcal{A}_{\bar{\mathcal{I}}})$ is:
\begin{equation}
    \mathcal{J}_{\text{bulk}} : \mathcal{A}_{\mathcal{I}} \longleftrightarrow \mathcal{A}_{\bar{\mathcal{I}}}, \qquad
\mathcal{J}_{\text{bulk}}(a) = J_{\text{bulk}} \, a \, J_{\text{bulk}}
\end{equation}
On the boundary, the theory is divided into two parts: the bath that absorbs Hawking radiation $\mathcal{R}$ and the CFT. Their operator algebras satisfy: $ \mathcal{A}_{\mathcal{R}} = \mathcal{A}_{\text{CFT}}'$. Let $|\Psi_{\text{bdy}}\rangle$ be the boundary state dual to the evaporating black hole geometry. The modular conjugation for the boundary is:
\begin{equation}
    \mathcal{J}_{\text{bdy}} : \mathcal{A}_{\mathcal{R}} \longleftrightarrow \mathcal{A}_{\text{CFT}}, \qquad
\mathcal{J}_{\text{bdy}}(O) = J_{\text{bdy}} \, O \, J_{\text{bdy}}.
\end{equation}
The reduced density matrices are defined as $\rho_{\mathcal{R}} = \operatorname{tr}_{\text{CFT}}(|\Psi_{\text{bdy}}\rangle\langle\Psi_{\text{bdy}}|)$ and $\rho_{\text{CFT}} = \operatorname{tr}_{\mathcal{R}}(|\Psi_{\text{bdy}}\rangle\langle\Psi_{\text{bdy}}|)$.

One important point here is that  $\bar{\mathcal{I}}$ is not simply the exterior — it includes both the exterior and a small region inside the horizon (although there are examples in which the island extends outside the horizon \cite{Geng:2020qvw}, here, we focus on the islands located in the interior of the black hole). Therefore, the standard HKLL map in AdS–Schwarzschild coordinates is insufficient. However, because $\bar{\mathcal{I}} = \mathcal{E}_{\text{CFT}}$, we can use the modular flow reconstruction of Faulkner and Lewkowycz \cite{Faulkner:2017} to define an embedding isometry $V_{\bar{\mathcal{I}}} : \mathcal{H}_{\bar{\mathcal{I}}} \to \mathcal{H}_{\text{CFT}}$ 
. For a bulk field $\phi(X)$ with $X \in \bar{\mathcal{I}}$, the reconstruction is:
\begin{widetext}
\begin{equation}
\begin{split}
    \Phi_{\text{CFT}}(X) = V_{\bar{\mathcal{I}}} \, \phi(X) \, V_{\bar{\mathcal{I}}}^\dagger
= &\int_{\text{CFT}} dx \int_{-\infty}^{\infty} ds \; f_{\Delta,s}(X|x) \; e^{i K_{\text{CFT}} s} \, \mathcal{O}(x) \, e^{-i K_{\text{CFT}} s} 
    \\
    =& \int_{\text{CFT}} dx \int_{-\infty}^{\infty} ds \; f_{\Delta,s}(X|x) \; \Delta_\psi ^{-i s} \, \mathcal{O}(x) \, \Delta_\psi ^{i s}
\end{split}
\end{equation}

\end{widetext}
where $K_{\text{CFT}}$  and $ \Delta_\psi$ are the modular Hamiltonian of the CFT state and the modular operator of the boundary state. 
The smearing function $f_{\Delta,s}(X|x)$ is defined via the modular Fourier transform and the Green's function method of \cite{Faulkner:2017}. Importantly, because the modular flow reconstruction works for any point in the entanglement wedge of the CFT, it remains valid for points in $\bar{\mathcal{I}}$ even if they lie behind the horizon, as long as they are in $\mathcal{E}_{\text{CFT}}$. For example, see \cite{Gao:2021tzr} for a calculation of correlation functions involving points behind the event horizon using modular flow.

Having this isometry, we define the Schrodinger-picture quantum channel
\begin{equation}\label{channel}
    \mathcal{N}_{\bar{\mathcal{I}}} : \mathcal{A}_{\bar{\mathcal{I}}} \longrightarrow \mathcal{A}_{\text{CFT}}, \qquad
\mathcal{N}_{\bar{\mathcal{I}}}(\cdot) = V_{\bar{\mathcal{I}}} \, (\cdot) \, V_{\bar{\mathcal{I}}}^\dagger.
\end{equation}

We now have all the ingredients to reconstruct island operators from the Hawking radiation. Starting from an island operator $\phi_{\mathcal{I}} \in \mathcal{A}_{\mathcal{I}}$, we proceed in three steps:

\begin{enumerate}
    \item Reflect to the complement via bulk modular conjugation:
    \[
    \phi_{\bar{\mathcal{I}}} = J_{\text{bulk}} \, \phi_{\mathcal{I}} \, J_{\text{bulk}} \in \mathcal{A}_{\bar{\mathcal{I}}}.
    \]
    
    \item Map to the CFT via $\psi$-dual of the  quantum channel \eqref{channel}:
    \[
    (\mathcal{N}^*)'_{\psi}(\phi_{\bar{\mathcal{I}}}) \in \mathcal{A}_{\text{CFT}}.
    \]
    
    \item  Reflect back via boundary modular conjugation:
  Apply $\mathcal{J}_{\text{bdy}}$ to make the expression manifestly in $\mathcal{A}_{\mathcal{R}}$.
\end{enumerate}

    \begin{tikzpicture}[>=stealth]
      \centering
    
    \node at ( 4.7,0) (L) {$\mathcal{A}_{\mathcal{R}}$};
    \node at ( 5,3) (3) {$\mathcal{A}_{\mathcal{I}}$};
    
    \node at ( 9,0)  (R){$\mathcal{A}_{CFT}$};
    \node at ( 9,3)  (1){$\mathcal{A}_{\bar{\mathcal{I}}}$};
    \draw [->, dashed] (9,2.6) -- (9,0.3);
    \node at ( 9.9 ,1.5)  {$(\mathcal{N}^*)'_{\psi}$};
    
    \draw [->] (R.west) -- (L.east);
    \draw [->] (3.east) -- (1.west);
    
    \node at ( 7,3.5)  {$\mathcal{J}_{bulk}$};
    \node at ( 7,-0.5)  {$\mathcal{J}_{bdy}$};
  
    \end{tikzpicture}

Putting these steps together, the \textbf{algebraic Petz map reconstruction of the island} is:
\begin{equation}
     \phi_{\mathcal{I}} \;\longrightarrow\; \mathcal{J}_{\text{bdy}} \circ (\mathcal{N}^*)'_\psi \circ \mathcal{J}_{\text{bulk}} (\phi_{\mathcal{I}}) \;\in\; \mathcal{A}_R \;
\end{equation}
Or in explicit notation:
\begin{equation}
   \Phi_{\mathcal{R}} = J_{\text{bdy}} \; (\mathcal{N}^*)'_\psi\!\left( J_{\text{bulk}} \, \phi_{\mathcal{I}} \, J_{\text{bulk}} \right) J_{\text{bdy}}. 
\end{equation}
The crucial point is that we never needed a direct map from the island to the boundary. Instead, we used modular conjugation to reflect into the complement, where a known map exists. 

\section{Free Scalar}

To make the reconstruction more concrete, consider a free scalar field theory in the bulk. To leading order in the AdS/CFT correspondence, bulk fields are free and can be expanded in modes. For any region $A$ on a Cauchy slice, the eigenfunctions of the modular Hamiltonian can be labeled by a frequency $\omega$ and a point $X_S$ on a co-dimension $2$ surface $S$ (which can be taken as the boundary of $A$) \cite{Faulkner:2017}. The corresponding mode operator satisfies $ [K_A, a_\omega(X_S)] = \omega \, a_\omega(X_S)$, where
\begin{equation}
    a_\omega(X_S) = \int ds \, e^{-i s \omega} \, e^{i K_A s} \, \phi(X_S) \, e^{-i K_A s}, \qquad \forall X_S \in S.
\end{equation}
The algebra of operators on $A$ is then spanned by these mode operators: $\mathcal{A}_A = \operatorname{span}\{ a_\omega(X_S) \}$.

We now apply this mode decomposition to the evaporating black hole after the Page time. Recall that the complement of the island, $\bar{\mathcal{I}}$, lies in the entanglement wedge of the entire CFT. Therefore, we can apply the modular flow reconstruction to this region. A natural choice for the co-dimension $2$ surface $S$ is the boundary of $\bar{\mathcal{I}}$ itself — which is precisely the Cauchy slice of the CFT. Hence, we write
\begin{equation}
    \mathcal{A}_{\bar{\mathcal{I}}} = \operatorname{span}\{ a_\omega(x) \mid \omega \in (-\infty, \infty),\; x \in \Sigma_{\text{CFT}}\},
\end{equation}
with the commutation relation $ [K_{\bar{\mathcal{I}}}, a_\omega(x)] = \omega \, a_\omega(x)$. 
The algebra of the island $\mathcal{I}$ is then obtained by acting with the bulk modular conjugation. Since $\mathcal{A}_{\mathcal{I}} = \mathcal{A}_{\bar{\mathcal{I}}}'$ and modular conjugation maps an algebra to its commutant, we have
\begin{equation}
    \mathcal{A}_{\mathcal{I}} = J_{\text{bulk}} \, \mathcal{A}_{\bar{\mathcal{I}}} \, J_{\text{bulk}} = \operatorname{span}\{ \tilde{a}_\omega(x) = J_{\text{bulk}} \, a_\omega(x) \, J_{\text{bulk}} \}.
\end{equation}
Thus, each island mode $\tilde{a}_\omega(x)$ is the modular conjugate of a CFT mode $a_\omega(x)$. Next, we use the embedding isometry $V_{\bar{\mathcal{I}}} : \mathcal{H}_{\bar{\mathcal{I}}} \to \mathcal{H}_{\text{CFT}}$ constructed via modular flow. Identifying the co-dimension $2$ surface with the CFT, the image of a bulk mode under this isometry is
\begin{equation}
    V_{\bar{\mathcal{I}}} \, a_\omega(x) \, V_{\bar{\mathcal{I}}}^\dagger = O_\omega(x) = \int ds \, e^{-i s \omega} \, \rho_{\text{CFT}}^{-i s} \, \mathcal{O}(x) \, \rho_{\text{CFT}}^{i s},
\end{equation}
where $\rho_{\text{CFT}} = \operatorname{tr}_{\mathcal{R}}(|\Psi_{\text{bdy}}\rangle\langle\Psi_{\text{bdy}}|)$ is the reduced density matrix of the CFT.

Now we apply the algebraic Petz map. Starting from an island mode $\tilde{a}_\omega(x) \in \mathcal{A}_{\mathcal{I}}$, the reconstruction proceeds as:
\begin{enumerate}
    \item Reflect to the complement via bulk modular conjugation (trivial here, since $\tilde{a}_\omega(x)$ is already defined as $J_{\text{bulk}} a_\omega(x) J_{\text{bulk}}$).
    \item Map to the CFT via the isometry: $a_\omega(x) \to O_\omega(x)$ through the $\psi$-dual map.
    \item Apply the boundary modular conjugation to obtain an operator in the bath algebra.
\end{enumerate}

The result is the \textbf{reconstructed island mirror operator}
\begin{equation}
    \tilde{O}_\omega(x) := J_{\text{bdy}} \, O_\omega(x) \, J_{\text{bdy}} = \mathcal{J}_{\text{bdy}} \circ (\mathcal{N}^*)'_\psi \circ \mathcal{J}_{\text{bulk}} (\tilde{a}_\omega(x)).
\end{equation}
This operator $\tilde{O}_\omega(x)$ lives in the commutant of the CFT — i.e., in the bath algebra $\mathcal{A}_{\mathcal{R}}$. In analogy with the Papadodimas–Raju proposal for equilibrium black holes \cite{Papadodimas:2012aq, Papadodimas:2013jku, Papadodimas:2015jra}, we refer to $\tilde{O}_\omega(x)$ as an \emph{island mirror operator}.

A central feature of our construction is the explicit emergence of state-dependence. Because the modular conjugations \(J_{\text{bulk}}\) and \(J_{\text{bdy}}\) are constructed relative to a specific reference state \(\ket{\psi}\), the reconstructed Hawking radiation operator \(\Phi_{\mathcal{R}} = J_{\text{bdy}} (\mathcal{N}^*)'_\psi(J_{\text{bulk}} \phi_{\mathcal{I}} J_{\text{bulk}}) J_{\text{bdy}}\) intrinsically depends on the background quantum state of the black hole. Rather than being an ad-hoc prescription, here state-dependence arises as a necessary consequence of Tomita--Takesaki modular theory in quantum error correction.

To connect more directly to the bath, we analyse the matrix elements of $\tilde{O}_\omega(x)$ in the boundary Hilbert space. The full boundary Hilbert space factorises as
$ \mathcal{H}_{\text{bdy}} = \mathcal{H}_{\text{CFT}} \otimes \mathcal{H}_{\mathcal{R}}$
and can also be represented as $\mathcal{H}_{\text{bdy}} = \overline{\mathcal{A}_{\text{CFT}} |\Psi_{\text{bdy}}\rangle}$. For any $a \in \mathcal{A}_{\text{CFT}}$, we associate a vector $|a\rangle = a |\Psi_{\text{bdy}}\rangle$. The matrix element of the island mirror operator between two such states is
\begin{widetext}
 \begin{equation}
    \begin{aligned}
\langle a | \tilde{O}_\omega(x) | b \rangle
&= \langle \Psi_{\text{bdy}} | a^\dagger \tilde{O}_\omega(x) b | \Psi_{\text{bdy}} \rangle \\
&= \langle \Psi_{\text{bdy}} | a^\dagger b \, \tilde{O}_\omega(x) | \Psi_{\text{bdy}} \rangle \qquad (\text{since } \tilde{O}_\omega(x) \text{ commutes with } \mathcal{A}_{\text{CFT}}) \\
&= \langle \Psi_{\text{bdy}} | a^\dagger b \, J_{\text{bdy}} \, O_\omega(x) \,J_{\text{bdy}} | \Psi_{\text{bdy}} \rangle \\
&= \langle \Psi_{\text{bdy}} | a^\dagger b \, \Delta_{\text{bdy}}^{1/2} \, O_{-\omega}(x) \,  | \Psi_{\text{bdy}} \rangle \qquad (\text{using } J \Delta^{1/2} = \Delta^{-1/2} J) \\
&= \langle \Psi_{\text{bdy}} | a^\dagger b \, \rho_{\text{CFT}}^{1/2} \, O_{-\omega}(x) \, \rho_{\text{CFT}}^{-1/2} | \Psi_{\text{bdy}} \rangle \\
&= \operatorname{tr}_{\text{CFT}} \Big( \rho_{\text{CFT}} \, a^\dagger b \, \rho_{\text{CFT}}^{1/2} \, O_{-\omega}(x) \, \rho_{\text{CFT}}^{-1/2} \Big) \\
&= \int ds \, e^{-i s \omega} \, \operatorname{tr}_{\text{CFT}} \Big( a^\dagger b \, \rho_{\text{CFT}}^{1/2 + i s} \, \mathcal{O}(x) \, \rho_{\text{CFT}}^{1/2 - i s} \Big).
\end{aligned}
\end{equation}   
\end{widetext}
Strictly speaking, the local QFT algebras appearing here are
generically of type III, for which a density matrix and a trace are
not intrinsically defined. The expressions above should therefore be
understood as shorthand for the corresponding algebraic constructions.
In the gravitational island setting, however, gravitational dressing
provides an important refinement: the algebra of appropriately dressed
island observables can be enlarged by the corresponding observer
degrees of freedom and identified, under the geometric modular-flow
conjecture, with a crossed-product algebra of type
$\mathrm{II}_{\infty}$~\cite{Geng:2025bcb}. Thus, trace and
density-matrix expressions can acquire an intrinsic meaning in the
gravitationally dressed algebra, rather than being merely a regulator
artifact.

To express this result purely in the bath, we choose an orthonormal basis for the CFT and the bath separately
\begin{equation}
    \mathcal{H}_{\text{CFT}} = \operatorname{span}\{ |i\rangle_{\text{CFT}} \}, \qquad
\mathcal{H}_{\mathcal{R}} = \operatorname{span}\{ |\tilde{j}\rangle_{\mathcal{R}} \}.
\end{equation}
where $  \ket{ i, \tilde{j}} = \sum_k \alpha_{i,j}^k \ket{a_k}$, where $ \ket{a_k} = a_k \ket{\Psi_{bdy}}$. One can obtain the matrix element of the mirror operator in the bath system that absorbed the Hawking radiation is
\begin{widetext}
\begin{equation}
\begin{aligned}
{}_{\mathcal R}\langle \tilde j'|
\tilde O_\omega(x)
|\tilde j\rangle_{\mathcal R}
&=
\sum_{k,k'}
\alpha_{ij}^{k}
\bigl(\alpha_{ij'}^{k'}\bigr)^*
\\
&\quad\times
\int ds\,e^{-is\omega}
\operatorname{tr}_{\mathrm{CFT}}
\Bigl(
a_{k'}^\dagger a_k\,
\rho_{\mathrm{CFT}}^{1/2+is}\,
\mathcal O(x)\,
\rho_{\mathrm{CFT}}^{1/2-is}
\Bigr).
\end{aligned}
\end{equation}
\end{widetext}

This expression gives the explicit action of the reconstructed island operator on the Hawking radiation state.
This is a formal expression for radiation matrix elements, conditional on the required modular data and the identification of the standard representation with the physical bath degrees of freedom

For the evaporating case, these island mirror operators provide the explicit map from the island to the Hawking radiation, confirming that the information in the island is accessible to the outside observer after the Page time.

\section{Discussion}

Understanding the operational decoding of Hawking radiation is a central challenge in quantum gravity. While the island proposal resolves the entropic aspect of the black-hole information paradox by producing a unitary Page curve, it does not explicitly specify how operators deep in the black-hole interior are encoded in the asymptotic Hawking radiation. In this work, we have addressed this question by constructing an operator-level holographic dictionary for black-hole islands using the algebraic Petz map.

The core physical insight behind our construction is that the reconstruction problem can be transferred to the commutant algebra. Although the island \(\mathcal{I}\) lacks a direct bulk-to-boundary reconstruction channel, its complement \(\bar{\mathcal{I}}\) lies entirely within the entanglement wedge of the asymptotic CFT after the Page time. By employing Tomita--Takesaki modular conjugation \(J_{\text{bulk}}\), island operators are reflected into this accessible complementary domain, transported to the boundary via the state-dependent dual channel \((\mathcal{N}^*)'_\psi\), and mapped into the radiation algebra \(\mathcal{A}_{\mathcal{R}}\) using the boundary modular conjugation \(J_{\text{bdy}}\). This bypasses the need for global bulk-to-boundary isometries or continuous modular flows.


Furthermore, we have shown that the reconstructed island modes precisely match the structure of Papadodimas-Raju mirror operators. In our algebraic formulation, the \emph{mirror} is naturally identified with the modular conjugation of the boundary theory. Since the algebraic Petz map transforms bulk interior modes into non-local boundary operators, the explicit spatial location of the quantum island \(\mathcal{I}\) can be probed directly through boundary operator growth and Krylov complexity~\cite{Vardian:2026yar}, providing a precise bridge between algebraic recovery and quantum complexity.

Looking forward, this algebraic framework opens several exciting avenues. First, while our construction relies on the leading-order $1/N$ expansion where bulk fields are free, understanding how this algebraic reflection incorporates non-perturbative gravitational corrections and quantum complexity will be crucial for a complete theory of decoding. It is worth noting that, in principle, one could employ the algebraic version of the twirled Petz recovery channel in the Heisenberg picture \cite{faulkner2022approximate,faulkner2022approximate2} to incorporate \(1/N\) corrections. However, a limitation of this approach is that the EWR based on modular flow, as formulated by Faulkner and Lewkowycz, is itself valid only in the large-\(N\) limit. However, the first subleading correction to this formula has been derived recently \cite{Shahmiri:2026jkx}, though its applicability is currently limited to spacetimes without black holes.

It would be useful to evaluate the present construction in the
free-fermion island models studied in
\cite{Chen:2019iro} and compare the resulting radiation
representatives with the explicit modular-flow reconstruction.
Such a comparison could clarify the relation between the two
constructions and the modular data required by each.

\section{Acknowledgments}

I  would like to thank A. A. Ardehali and K. Papadodimas for  useful discussions, and H. Arfaei for his valuable comments on the initial draft. Moreover, this work was supported by the "Tarh-e-Sobat" Grant No 11/157663. I also acknowledge the use of ChatGPT (OpenAI), Claude and Gemini as  editorial assistants in improving the clarity and presentation of the manuscript.

\bibliographystyle{apsrev4-2} 
\bibliography{refs.bib} 

@article{Kaushal:2025wbn,
    author = "Kaushal, Anurag and Prabhakar, Naveen S. and Wadia, Spenta R.",
    title = "{Gauge invariant Hamiltonian evolution across the black hole horizon in asymptotically AdS spacetimes}",
    eprint = "2510.21920",
    archivePrefix = "arXiv",
    primaryClass = "hep-th",
    doi = "10.1103/gz2s-mqk5",
    journal = "Phys. Rev. D",
    volume = "113",
    number = "10",
    pages = "106019",
    year = "2026"
}

@article{Chen:2019iro,
    author = "Chen, Yiming",
    title = "{Pulling Out the Island with Modular Flow}",
    eprint = "1912.02210",
    archivePrefix = "arXiv",
    primaryClass = "hep-th",
    doi = "10.1007/JHEP03(2020)033",
    journal = "JHEP",
    volume = "03",
    pages = "033",
    year = "2020"
}

@article{jafferis2016relative,
  title={Relative entropy equals bulk relative entropy},
  author={Jafferis, Daniel L and Lewkowycz, Aitor and Maldacena, Juan and Suh, S Josephine},
  journal={Journal of High Energy Physics},
  volume={2016},
  number={6},
  pages={1--20},
  year={2016},
  publisher={Springer}
}

@article{cotler2019entanglement,
  title={Entanglement wedge reconstruction via universal recovery channels},
  author={Cotler, Jordan and Hayden, Patrick and Penington, Geoffrey and Salton, Grant and Swingle, Brian and Walter, Michael},
  journal={Physical Review X},
  volume={9},
  number={3},
  pages={031011},
  year={2019},
  publisher={APS}
}

@article{Faulkner:2017,
  author       = {Faulkner, Thomas and Lewkowycz, Aitor},
  title        = {Bulk locality from modular flow},
  journal      = {JHEP},
  volume       = {07},
  year         = {2017},
  pages        = {151},
  doi          = {10.1007/JHEP07(2017)151},
  eprint       = {1704.05464},
  archivePrefix= {arXiv},
  primaryClass = {hep-th}
}

@article{chen2020entanglement,
  title={Entanglement wedge reconstruction using the Petz map},
  author={Chen, Chi-Fang and Penington, Geoffrey and Salton, Grant},
  journal={Journal of High Energy Physics},
  volume={2020},
  number={1},
  pages={1--14},
  year={2020},
  publisher={Springer}
}

@article{hamilton2006local,
  title={Local bulk operators in AdS/CFT correspondence: A boundary view of horizons and locality},
  author={Hamilton, Alex and Kabat, Daniel and Lifschytz, Gilad and Lowe, David A},
  journal={Physical Review D},
  volume={73},
  number={8},
  pages={086003},
  year={2006},
  publisher={APS}
}

@article{hamilton2006holographic,
  title={Holographic representation of local bulk operators},
  author={Hamilton, Alex and Kabat, Daniel and Lifschytz, Gilad and Lowe, David A},
  journal={Physical Review D},
  volume={74},
  number={6},
  pages={066009},
  year={2006},
  publisher={APS}
}

@article{hamilton2007local,
  title={Local bulk operators in AdS/CFT correspondence: A holographic description of the black hole interior},
  author={Hamilton, Alex and Kabat, Daniel and Lifschytz, Gilad and Lowe, David A},
  journal={Physical Review D},
  volume={75},
  number={10},
  pages={106001},
  year={2007},
  publisher={APS}
}

@article{hamilton2008local,
  title={Local bulk operators in AdS/CFT and the fate of the BTZ singularity},
  author={Hamilton, Alex and Kabat, DN and Lifschytz, Gilad and Lowe, David A},
  journal={AMS/IP Stud. Adv. Math},
  volume={44},
  pages={85},
  year={2008}
}

@article{Papadodimas:2013jku,
    author = "Papadodimas, Kyriakos and Raju, Suvrat",
    title = "{State-Dependent Bulk-Boundary Maps and Black Hole Complementarity}",
    eprint = "1310.6335",
    archivePrefix = "arXiv",
    primaryClass = "hep-th",
    reportNumber = "ICTS-2013-21, ICTS/2013/21",
    doi = "10.1103/PhysRevD.89.086010",
    journal = "Phys. Rev. D",
    volume = "89",
    number = "8",
    pages = "086010",
    year = "2014"
}

@article{Vardian:2026yar,
    author = "Vardian, Niloofar",
    title = "{Modular Krylov Complexity as a Boundary Probe of Area Operator and Entanglement Islands}",
    eprint = "2602.02675",
    archivePrefix = "arXiv",
    primaryClass = "hep-th",
    month = "2",
    year = "2026",
    journal=""
}

@article{Bahiru:2023zlc,
    author = "Bahiru, Eyoab and Belin, Alexandre and Papadodimas, Kyriakos and Sarosi, Gabor and Vardian, Niloofar",
    title = "{Holography and localization of information in quantum gravity}",
    eprint = "2301.08753",
    archivePrefix = "arXiv",
    primaryClass = "hep-th",
    reportNumber = "CERN-TH-2023-003",
    doi = "10.1007/JHEP05(2024)261",
    journal = "JHEP",
    volume = "05",
    pages = "261",
    year = "2024"
}

@article{Bahiru:2022oas,
    author = "Bahiru, Eyoab and Belin, Alexandre and Papadodimas, Kyriakos and Sarosi, Gabor and Vardian, Niloofar",
    title = "{State-dressed local operators in the AdS/CFT correspondence}",
    eprint = "2209.06845",
    archivePrefix = "arXiv",
    primaryClass = "hep-th",
    reportNumber = "CERN-TH-2022-145",
    doi = "10.1103/PhysRevD.108.086035",
    journal = "Phys. Rev. D",
    volume = "108",
    number = "8",
    pages = "086035",
    year = "2023"
}

@article{Papadodimas:2015jra,
    author = "Papadodimas, Kyriakos and Raju, Suvrat",
    title = "{Remarks on the necessity and implications of state-dependence in the black hole interior}",
    eprint = "1503.08825",
    archivePrefix = "arXiv",
    primaryClass = "hep-th",
    doi = "10.1103/PhysRevD.93.084049",
    journal = "Phys. Rev. D",
    volume = "93",
    number = "8",
    pages = "084049",
    year = "2016"
}

@article{Papadodimas:2012aq,
    author = "Papadodimas, Kyriakos and Raju, Suvrat",
    title = "{An Infalling Observer in AdS/CFT}",
    eprint = "1211.6767",
    archivePrefix = "arXiv",
    primaryClass = "hep-th",
    reportNumber = "HRI-ST-1107, ICTS-2012-10",
    doi = "10.1007/JHEP10(2013)212",
    journal = "JHEP",
    volume = "10",
    pages = "212",
    year = "2013"
}

@article{Almheiri:2014lwa,
    author = "Almheiri, Ahmed and Dong, Xi and Harlow, Daniel",
    title = "{Bulk Locality and Quantum Error Correction in AdS/CFT}",
    eprint = "1411.7041",
    archivePrefix = "arXiv",
    primaryClass = "hep-th",
    reportNumber = "SU-ITP-14-30, SU-ITP-14/30",
    doi = "10.1007/JHEP04(2015)163",
    journal = "JHEP",
    volume = "04",
    pages = "163",
    year = "2015"
}

@article{Chen:2019gbt,
    author = "Chen, Chi-Fang and Penington, Geoffrey and Salton, Grant",
    title = "{Entanglement Wedge Reconstruction using the Petz Map}",
    eprint = "1902.02844",
    archivePrefix = "arXiv",
    primaryClass = "hep-th",
    doi = "10.1007/JHEP01(2020)168",
    journal = "JHEP",
    volume = "01",
    pages = "168",
    year = "2020"
}

@article{Bahiru:2022ukn,
    author = "Bahiru, Eyoab and Vardian, Niloofar",
    title = "{Explicit reconstruction of the entanglement wedge via the Petz map}",
    eprint = "2210.00602",
    archivePrefix = "arXiv",
    primaryClass = "hep-th",
    doi = "10.1007/JHEP07(2023)025",
    journal = "JHEP",
    volume = "07",
    pages = "025",
    year = "2023"
}

@article{Vardian:2023fce,
    author = "Vardian, Niloofar",
    title = "{Black hole interior Petz map reconstruction and Papadodimas-Raju proposal}",
    eprint = "2307.01858",
    archivePrefix = "arXiv",
    primaryClass = "hep-th",
    doi = "10.1007/JHEP10(2023)024",
    journal = "JHEP",
    volume = "10",
    pages = "024",
    year = "2023"
}

@article{Penington:2019npb,
    author = "Penington, Geoffrey",
    title = "{Entanglement Wedge Reconstruction and the Information Paradox}",
    eprint = "1905.08255",
    archivePrefix = "arXiv",
    primaryClass = "hep-th",
    doi = "10.1007/JHEP09(2020)002",
    journal = "JHEP",
    volume = "09",
    pages = "002",
    year = "2020"
}

@article{Almheiri:2019psf,
    author = "Almheiri, Ahmed and Engelhardt, Netta and Marolf, Donald and Maxfield, Henry",
    title = "{The entropy of bulk quantum fields and the entanglement wedge of an evaporating black hole}",
    eprint = "1905.08762",
    archivePrefix = "arXiv",
    primaryClass = "hep-th",
    doi = "10.1007/JHEP12(2019)063",
    journal = "JHEP",
    volume = "12",
    pages = "063",
    year = "2019"
}

@article{page1993information,
  title={Information in black hole radiation},
  author={Page, Don N},
  journal={Physical review letters},
  volume={71},
  number={23},
  pages={3743},
  year={1993},
  publisher={APS}
}

@article{Bena:1999jv,
    author = "Bena, Iosif",
    title = "{On the construction of local fields in the bulk of AdS(5) and other spaces}",
    eprint = "hep-th/9905186",
    archivePrefix = "arXiv",
    doi = "10.1103/PhysRevD.62.066007",
    journal = "Phys. Rev. D",
    volume = "62",
    pages = "066007",
    year = "2000"
}

@article{Cotler:2017erl,
    author = "Cotler, Jordan and Hayden, Patrick and Penington, Geoffrey and Salton, Grant and Swingle, Brian and Walter, Michael",
    title = "{Entanglement Wedge Reconstruction via Universal Recovery Channels}",
    eprint = "1704.05839",
    archivePrefix = "arXiv",
    primaryClass = "hep-th",
    doi = "10.1103/PhysRevX.9.031011",
    journal = "Phys. Rev. X",
    volume = "9",
    number = "3",
    pages = "031011",
    year = "2019"
}

@article{Hayden:2018khn,
    author = "Hayden, Patrick and Penington, Geoffrey",
    title = "{Learning the Alpha-bits of Black Holes}",
    eprint = "1807.06041",
    archivePrefix = "arXiv",
    primaryClass = "hep-th",
    doi = "10.1007/JHEP12(2019)007",
    journal = "JHEP",
    volume = "12",
    pages = "007",
    year = "2019"
}

@article{Gao:2021tzr,
    author = "Gao, Ping and Lamprou, Lampros",
    title = "{Seeing behind black hole horizons in SYK}",
    eprint = "2111.14010",
    archivePrefix = "arXiv",
    primaryClass = "hep-th",
    doi = "10.1007/JHEP06(2022)143",
    journal = "JHEP",
    volume = "06",
    pages = "143",
    year = "2022"
}

@article{furuya2020real,
  title={Real-space renormalization, error correction and conditional expectations},
  author={Furuya, Keiichiro and Lashkari, Nima and Ouseph, Shoy},
  journal={arXiv preprint arXiv:2012.14001},
  year={2020}
}

@article{faulkner2022approximate,
  title={Approximate Recovery and Relative Entropy I: General von Neumann Subalgebras},
  author={Faulkner, Thomas and Hollands, Stefan and Swingle, Brian and Wang, Yixu},
  journal={Communications in Mathematical Physics},
  volume={389},
  number={1},
  pages={349--397},
  year={2022},
  publisher={Springer}
}

@article{faulkner2022approximate2,
  title={Approximate recoverability and relative entropy II: 2-positive channels of general von Neumann algebras},
  author={Faulkner, Thomas and Hollands, Stefan},
  journal={Letters in Mathematical Physics},
  volume={112},
  number={2},
  pages={1--24},
  year={2022},
  publisher={Springer}
}

@article{Shahmiri:2026jkx,
    author = "Shahmiri, Arash Alipour and Sharifian, Maryam and Vardian, Niloofar",
    title = "{Entanglement Wedge Reconstruction Beyond the Large N Limit via the Twirled Petz Map}",
    eprint = "2609.07866",
    archivePrefix = "arXiv",
    primaryClass = "hep-th",
    month = "9",
    year = "2026",
    journal=""
}

@article{Geng:2020qvw,
    author = "Geng, Hao and Karch, Andreas",
    title = "{Massive islands}",
    eprint = "2006.02438",
    archivePrefix = "arXiv",
    primaryClass = "hep-th",
    doi = "10.1007/JHEP09(2020)121",
    journal = "JHEP",
    volume = "09",
    pages = "121",
    year = "2020"
}

@article{Geng:2025bcb,
    author = "Geng, Hao and Jiang, Yikun and Xu, Jiuci",
    title = "{Algebras, entanglement islands, and observers}",
    eprint = "2506.12127",
    archivePrefix = "arXiv",
    primaryClass = "hep-th",
    doi = "10.1088/1361-6382/ae811e",
    journal = "Class. Quant. Grav.",
    volume = "43",
    number = "13",
    pages = "135012",
    year = "2026"
}

@article{Geng:2024dbl,
    author = "Geng, Hao",
    title = "{Quantum rods and clock in a gravitational universe}",
    eprint = "2412.03636",
    archivePrefix = "arXiv",
    primaryClass = "hep-th",
    doi = "10.1103/kpqb-jrc8",
    journal = "Phys. Rev. D",
    volume = "112",
    number = "2",
    pages = "026009",
    year = "2025"
}

@article{Geng:2025rov,
    author = "Geng, Hao",
    title = "{The mechanism behind the information encoding for islands}",
    eprint = "2502.08703",
    archivePrefix = "arXiv",
    primaryClass = "hep-th",
    doi = "10.1007/JHEP03(2026)037",
    journal = "JHEP",
    volume = "03",
    pages = "037",
    year = "2026"
}

@article{Geng:2025gns,
    author = "Geng, Hao and Huertas, Jes{\'u}s and Karch, Andreas and Randall, Lisa and Thomas, Dawson",
    title = "{Wet Hair: Global Symmetries in Entanglement Islands}",
    eprint = "2512.11025",
    archivePrefix = "arXiv",
    primaryClass = "hep-th",
    month = "12",
    year = "2025",
    journal=""
    }

@article{Geng:2025byh,
    author = "Geng, Hao",
    title = "{Making the case for massive islands}",
    eprint = "2509.22775",
    archivePrefix = "arXiv",
    primaryClass = "hep-th",
    doi = "10.1007/JHEP08(2026)214",
    journal = "JHEP",
    volume = "08",
    pages = "214",
    year = "2026"
}

@article{Geng:2025gqu,
    author = "Geng, Hao and Jafferis, Daniel and Shrivastava, Pushkal and Tata, Neeraj",
    title = "{The Fate of Information Localizability and Holography in Quantum Gravity}",
    eprint = "2512.18912",
    archivePrefix = "arXiv",
    primaryClass = "hep-th",
    month = "12",
    year = "2025",
    journal=""
}

@article{Geng:2026asi,
    author = "Geng, Hao and Karch, Andreas and Perez-Pardavila, Carlos and Raju, Suvrat and Randall, Lisa and Riojas, Marcos",
    title = "{Seeing Page Curves and Islands with Blinders On}",
    eprint = "2602.06543",
    archivePrefix = "arXiv",
    primaryClass = "hep-th",
    month = "2",
    year = "2026",
      journal=""
}

@misc{antonini2025apologia,
      title={An apologia for islands},
      author={Stefano Antonini and Chang-Han Chen and Henry Maxfield and Geoff Penington},
      year={2025},
      eprint={2506.04311},
      archivePrefix={arXiv},
      doi={10.48550/arXiv.2506.04311},
      url={https://arxiv.org/abs/2506.04311},
}

@misc{elsayed2026observablealgebrajtdesitter,
      title={Toward an Observable Algebra for JT-de Sitter Space: Gap Protection and Modular Dressings}, 
      author={Hassan ElSayed},
      year={2026},
      eprint={2607.26194},
      archivePrefix={arXiv},
      primaryClass={hep-th},
      url={https://arxiv.org/abs/2607.26194}, 
}

\clearpage
\onecolumngrid

\appendix

\section{Algebraic Petz map}\label{A}

First, we briefly recall the relevant notions from the theory of von Neumann algebras. A von Neumann algebra $\mathcal{A}$ in its standard form is defined as a quadruple $(\mathcal{A}, \mathcal{H}_A, \mathcal{J}_A, \mathcal{P}_A)$, where:
\begin{itemize}
    \item $\mathcal{A}$ acts on the Hilbert space $\mathcal{H}_A$.
    \item $\mathcal{J}_A = J_A (.) J_A$ is an anti-linear, unitary involution called the \emph{modular conjugation}.
    \item $\mathcal{P}_A$ is the \emph{natural cone}, which is invariant under $\mathcal{J}_A$.
\end{itemize}
Here, the key object for us is the modular conjugation $J_A$. It exists whenever we have a von Neumann algebra and a cyclic-separating state. Its most important property is that it maps an algebra to its commutant.

Let $\mathcal{A}$ be an algebra of operators acting on a Hilbert space $\mathcal{H}$, and let $|\Psi\rangle$ be a \emph{cyclic and separating} state for $\mathcal{A}$. Cyclic means that acting with operators from $\mathcal{A}$ on $|\Psi\rangle$ fills the whole Hilbert space. Separating means that if $a|\Psi\rangle = 0$ for some $a \in \mathcal{A}$, then $a = 0$. In AdS/CFT, vacuum states typically have these properties for local subalgebras.

For such a state, Tomita–Takesaki theory tells us that there exists an anti-linear operator $J_A$ called the \emph{modular conjugation} satisfying:
\begin{equation}
  J_A \, \mathcal{A} \, J_A = \mathcal{A}',  
\end{equation}
where $\mathcal{A}'$ is the \emph{commutant} of $\mathcal{A}$ — the set of all operators that commute with every operator in $\mathcal{A}$. In other words, $J_A$ provides an anti-isomorphism between an algebra and its commutant.

In AdS/CFT:
\begin{itemize}
    \item On the boundary: For a region $A$ and its complement $\bar{A}$, we have $(\mathcal{A}_A)' = \mathcal{A}_{\bar{A}}$. The modular conjugation $J_{\text{bdy}}$ maps operators in $A$ to operators in $\bar{A}$.
    \item In the bulk: For an entanglement wedge $a$ and its complement $\bar{a}$, we have $(\mathcal{A}_a)' = \mathcal{A}_{\bar{a}}$. The bulk modular conjugation $J_{\text{bulk}}$ maps operators in $a$ to operators in $\bar{a}$.
\end{itemize}
Thus, modular conjugation provides an algebraic way to \emph{reflect} operators across the RT surface, from a region to its complement. 

Now consider two von Neumann algebras $(\mathcal{A}, \mathcal{H}_A, \mathcal{J}_A, \mathcal{P}_A)$ and $(\mathcal{B}, \mathcal{H}_B, \mathcal{J}_B, \mathcal{P}_B)$. Let $\mathcal{T}: \mathcal{B} \to \mathcal{A}$ be a unital completely positive map between the algebras. In the context of QEC, this map corresponds to the Heisenberg-picture version of a quantum channel.

Choose a faithful state $\rho_A$ on $\mathcal{A}$, and define $\rho_B = \mathcal{T}^*(\rho_A)$ on $\mathcal{B}$. If the relative entropy is preserved for all states — i.e.,
\begin{equation}
    D(\rho \| \sigma) = D(\mathcal{T}^*(\rho) \| \mathcal{T}^*(\sigma)) \qquad \forall \rho,\sigma \in \mathcal{S}(\mathcal{A}),
\end{equation}
then there exists a recovery map $\mathcal{T}^P_\psi : \mathcal{A} \to \mathcal{B}$ such that $\mathcal{T} \circ \mathcal{T}^P_\psi$ acts as the identity on $\mathcal{H}_A$. This recovery map is given by
\begin{equation}
 \mathcal{T}^P_\psi(\cdot) = \mathcal{J}_B \circ \mathcal{T}'_\psi \circ \mathcal{J}_A (\cdot),   
\end{equation}
where $\mathcal{T}'_\psi : \mathcal{A}' \to \mathcal{B}'$ is the \emph{$\psi$-dual map} acting on the commutants.
To define $\mathcal{T}'_\psi$, we use the GNS representation. Let $|\Psi_A\rangle \in \mathcal{P}_A$ and $|\Psi_B\rangle \in \mathcal{P}_B$ be the unique vector representatives of the states $\rho_A$ and $\rho_B$. The GNS Hilbert spaces satisfy $\mathcal{H}_A = \overline{\mathcal{A}' |\Psi_A\rangle}$ and $\mathcal{H}_B = \overline{\mathcal{B}' |\Psi_B\rangle}$. Given the superoperator $\mathcal{T}$ and its corresponding operator $T: \mathcal{H}_B \to \mathcal{H}_A$ between GNS Hilbert spaces, the $\psi$-dual map is defined by
\begin{equation}
  \langle a' | \mathcal{T}(b) \rangle_{\Psi_A} = \langle \mathcal{T}'_\rho(a') | b \rangle_{\Psi_B}, \qquad \forall a' \in \mathcal{A}', \; b \in \mathcal{B}.  
\end{equation}

In the case of matrix algebras (finite-dimensional systems), this reduces to the familiar expression
\begin{equation}
   \operatorname{tr}\!\big( \rho_A \, a^\dagger \, \mathcal{T}(b) \big) = \operatorname{tr}\!\big( \rho_B \, \mathcal{T}^*_\rho(a^\dagger) \, b \big), 
\end{equation}
 and $\mathcal{T}'_\psi$ becomes equivalent to the dual of the Petz recovery channel.
Indeed, the $\psi$-dual map $\mathcal{T}'_\psi$  acts on the commutant algebras and reverses the action of the original map \cite{furuya2020real, faulkner2022approximate,faulkner2022approximate2}.

We now apply this algebraic framework to entanglement wedge reconstruction. Consider a semiclassical CFT state $|\Psi_g\rangle$ dual to a bulk geometry $g_{\mu\nu}$, and let $|\psi_g\rangle$ be the corresponding bulk quantum state. Let $A$ be a boundary region with entanglement wedge $a = \mathcal{E}_A$. The map of interest is
\begin{equation}
 \mathcal{N} : \mathcal{A}_a \longrightarrow \mathcal{A}_A,   
\end{equation}
which sends bulk operators in the entanglement wedge to boundary operators on $A$.

Because the regions and their complements satisfy
\begin{equation}
 (\mathcal{A}_A)' = \mathcal{A}_{\bar{A}}, \qquad (\mathcal{A}_a)' = \mathcal{A}_{\bar{a}},   
\end{equation}
and assuming $|\Psi_g\rangle$ and $|\psi_g\rangle$ are cyclic and separating for their respective algebras, we have the modular conjugations:
\begin{equation}
\begin{aligned}
\mathcal{J}_{\text{bulk}} &: \; \mathcal{A}_a \longleftrightarrow \mathcal{A}_{\bar{a}}, \qquad 
\mathcal{J}_{\text{bulk}}(a) = J_{\text{bulk}} \, a \, J_{\text{bulk}}, \\
\mathcal{J}_{\text{CFT}} &: \; \mathcal{A}_A \longleftrightarrow \mathcal{A}_{\bar{A}}, \qquad 
\mathcal{J}_{\text{CFT}}(O) = J_{\text{CFT}} \, O \, J_{\text{CFT}}.
\end{aligned}    
\end{equation}

The reconstruction proceeds in three steps:

\begin{enumerate}
    \item \textbf{Reflect to the complement via bulk modular conjugation:}
    Start with a bulk operator $\phi_a \in \mathcal{A}_a$ in the entanglement wedge. Apply bulk modular conjugation to obtain an operator in the commutant:
    \begin{equation}
      \phi_{\bar{a}} = J_{\text{bulk}} \, \phi_a \, J_{\text{bulk}} \in \mathcal{A}_{\bar{a}}.  
    \end{equation}

    \item \textbf{Apply the $\psi$-dual recovery map:}
    Following the same logic as \cite{Vardian:2023fce}, send this operator to the boundary commutant using the $\psi$-dual map:
\begin{equation}
        (\mathcal{N}^*)'_\psi(\phi_{\bar{a}}) \in \mathcal{A}'_A = \mathcal{A}_{\bar{A}}.
\end{equation}
This formulation demonstrates that whenever a direct channel \(\mathcal{N}_{a} : \mathcal{A}_{a} \to \mathcal{A}_A\) lacks an explicit isometry, the existence of an accessible complement channel \(\mathcal{N}_{\bar{a}}\) alongside cyclic-separating reference states can be mathematically sufficient to rigorously reconstruct \(\mathcal{A}_{a}\) observables.

    \item \textbf{Reflect back to the original region via boundary modular conjugation:}
    Finally, apply boundary modular conjugation to obtain an operator supported on $A$:
    \begin{equation}
        \Phi_A = J_{\text{bdy}} \;  (\mathcal{N}^*)'_\psi(\phi_{\bar{a}}) \; J_{\text{bdy}} \in \mathcal{A}_A.
    \end{equation}
    
\end{enumerate}

Putting these steps together, we obtain the \textbf{algebraic Petz map}:
\begin{equation}
    \boxed{\; \Phi_A = \mathcal{J}_{\text{bdy}} \;\circ\;  (\mathcal{N}^*)'_\psi \;\circ\; \mathcal{J}_{\text{bulk}} \;(\phi_a) \; } .
\end{equation}

In explicit notation:
\begin{equation}
    \label{algebraic}
\Phi_A = J_{\text{bdy}} \;  (\mathcal{N}^*)'_\psi\!\left( J_{\text{bulk}} \, \phi_a \, J_{\text{bulk}} \right) J_{\text{bdy}}.
\end{equation} 
For readers familiar with modular theory, the algebraic Petz map can also be understood as the solution to a relation involving the KMS inner product. If $|\Psi\rangle$ is cyclic and separating for a von Neumann algebra $\mathcal{A}$, the KMS inner product is defined as
\begin{equation}
    \langle a_1 | a_2 \rangle_{\text{KMS}, \Psi} = \langle \mathcal{J}_\Psi(a_1^\dagger) | a_2 \rangle_\Psi = \langle \Psi | a_1^\dagger \Delta_\Psi^{1/2} a_2 | \Psi \rangle,
\end{equation}

where $\Delta_\Psi$ is the modular operator. The algebraic Petz map is then the unique map satisfying
\begin{equation}
    \langle a | \mathcal{T}(b) \rangle_{\Psi_A} = \langle \mathcal{T}^P_\rho(a) | b \rangle_{\Psi_B}
\end{equation}
with respect to this inner product \cite{furuya2020real, faulkner2022approximate,faulkner2022approximate2}. This formulation highlights the close connection between Petz recovery and modular theory, but the essential physics is already captured by the three-step procedure described above.

The main advantage of the algebraic Petz map for black hole interior reconstruction is that it shifts the reconstruction problem from the algebras of the regions themselves to their commutants, where the necessary maps might be known. In the black hole setting, we do not know a global isometry from the full bulk (including the interior) to the boundary. However, we can have  well-defined isometries  for the complementary part to the boundary theory. The interior algebra is not directly accessible via the bulk-to-boundary map. But using the bulk modular conjugation $J_{\text{bulk}}$, we can map an interior operator $\phi_{\text{int}} \in \mathcal{A}_{\text{int}}$ to an operator in the commutant. This reflection places the operator into a domain where a map in principle can be defined. The $\psi$-dual map $(\mathcal{N}^*)'_\psi$ then acts on the commutant  and sends it to the boundary commutant. Finally, the boundary modular conjugation $J_{\text{bdy}}$ maps this back to the boundary algebra of interest. The crucial point is that the algebraic Petz map never requires a direct map from the interior to the boundary. Instead, it cleverly uses modular conjugation to reflect interior operators into the commutant which is mostly  the exterior, where the exterior channel already provides a controlled recovery map. This bypasses the need for a global isometry and is why the algebraic Petz map succeeds where the ordinary Petz map fails: it reconstructs the black hole interior using only the known exterior dynamics and the universal modular structure that exists for any cyclic-separating state.

\end{document}